%% file: vignali.tex
\documentclass[useAMS,galley]{mn2e}
\usepackage{graphicx}
\usepackage{times,mathptm}
\usepackage{lineno}
\newcommand{\ltsima}{$\; \buildrel < \over \sim \;$}
\newcommand{\simlt}{\lower.5ex\hbox{\ltsima}}
\newcommand{\gtsima}{$\; \buildrel > \over \sim \;$}
\newcommand{\simgt}{\lower.5ex\hbox{\gtsima}}

\newcommand{\cgs}{ ${\rm erg~cm}^{-2}~{\rm s}^{-1}$} 
\newcommand{\lum}{\rm erg~s$^{-1}$}

\def\lesssim{\mathrel{\hbox{\rlap{\hbox{\lower4pt\hbox{$\sim$}}}\hbox{$<$}}}}
\def\gtrsim{\mathrel{\hbox{\rlap{\hbox{\lower4pt\hbox{$\sim$}}}\hbox{$>$}}}}

\def\arcsec{\hbox{$^{\prime\prime}$}}
\def\mjy{\hbox{$\mu$Jy}}

\def\ebv{\hbox{$E(B\!-\!V)$}}

\def\aox{$\alpha_{\rm ox}$}

\def\ab1450{$AB_{1450(1+z)}$}

\def\xray{\hbox{X-ray}}

\def\mgii{\hbox{Mg\ {\sc ii}}}
\def\civ{\hbox{C\ {\sc iv}}}
\def\heii{\hbox{He\ {\sc ii}}}

\def\gd{GD~158\_19}

\def\msun{M$_{\odot}$}
\def\sfr{M$_{\odot}$~yr$^{-1}$}
\def\edd_ratio{$\log\ L_{\rm bol}/L_{\rm Edd}$}

\def\chandra{{\it Chandra\/}}

\def\heao1{{\it HEAO-1\/}}

\def\spitzer{{\it Spitzer\/}}
\def\scuba{{\it SCUBA\/}}

\def\xmm{{XMM-{\it Newton\/}}}

\def\aj{AJ}
\def\araa{ARA\&A}
\def\apj{ApJ}
\def\apjl{ApJ}
\def\apjs{ApJS}
\def\apss{Ap\&SS}
\def\aap{A\&A}
\def\mnras{MNRAS}
\def\pasp{PASP} 
\def\nat{Nature}
\def\sci{Science}
\def\physrep{Phys.~Rep.} 
\title[THE BROAD-BAND VIEW OF A z=2 TYPE~2 QSO]
{The HELLAS2XMM Survey. XII. The infrared/sub-millimeter view of an \xray\ selected Type~2 quasar at $\mathbf{z\approx2}$}
\author[C. Vignali et al.]
{
C. Vignali$^{1,2}$\thanks{E-mail: cristian.vignali@unibo.it (CV).
},
F. Pozzi$^{1}$, 
J. Fritz$^{3}$, 
A. Comastri$^{2}$,
C. Gruppioni$^{2}$, 
E. Bellocchi$^{2,1}$,
\newauthor
F. Fiore$^{4}$, 
M. Brusa$^{5}$, 
R. Maiolino$^{4}$, 
M. Mignoli$^{2}$, 
F. La Franca$^{6}$, 
L. Pozzetti$^{2}$, 
G. Zamorani$^{2}$ 
\newauthor
and A. Merloni$^{7,5}$ \\ \\
$^{1}$ Dipartimento di Astronomia, Universit\`a degli Studi di Bologna, 
Via Ranzani 1, I--40127 Bologna, Italy \\
$^{2}$ INAF -- Osservatorio Astronomico di Bologna, Via Ranzani 1, 
I--40127 Bologna, Italy \\
$^{3}$ INAF -- Osservatorio Astronomico di Padova, Vicolo dell'Osservatorio 5, I--35122 Padova, Italy \\
$^{4}$ INAF -- Osservatorio Astronomico di Roma, Via Frascati 33, I--00040 Monteporzio, Roma, Italy \\
$^{5}$ Max-Planck-Institut f\"ur Extraterrestrische Physik (MPE), Giessenbachstr. 1, 85748 Garching, Germany \\
$^{6}$ Dipartimento di Fisica, Universit\`a di Roma Tre, Via della Vasca Navale 84, I--00146 Roma, Italy \\
$^{7}$ Excellence Cluster Universe, Boltzmannstr. 2, D-85748 Garching, Germany
}

\begin{document}
%\linenumbers*

\date{Accepted 2009 February 20. Received 2009 February 18; in original form 2008 October 13}

\pagerange{\pageref{firstpage}--\pageref{lastpage}} \pubyear{2008}

\maketitle

\label{firstpage}

\begin{abstract}
We present multi-wavelength 
observations (from optical to sub-millimeter, including \spitzer\ and \scuba) 
of H2XMMJ~003357.2$-$120038 (also \gd), an \xray\ selected, luminous narrow-line 
(Type~2) quasar at $z$=1.957 selected from the HELLAS2XMM survey. 
Its broad-band properties can be reasonably well modeled assuming three components: 
a stellar component to account for the optical and near-IR emission, 
an AGN component (i.e., dust heated by an accreting active nucleus), dominant in the mid-IR, 
with an optical depth at 9.7~\micron\ along the line of sight (close to the 
equatorial plane of the obscuring matter) of $\tau(9.7)=1$ 
and a full covering angle of the reprocessing matter (torus) of 140~degrees, 
and a far-IR starburst component (i.e., dust heated by star formation) 
to reproduce the wide bump observed longward of 70~\micron. 
The derived star-formation rate is $\approx$~1500~\sfr. 
The overall modeling indicates that \gd\ is a high-redshift \xray\ luminous, 
obscured quasar with coeval powerful AGN activity and intense star formation. 
It is probably caught before the process of expelling the obscuring gas has started, 
thus quenching the star formation. 
\end{abstract}


\section{Introduction}
\label{introduction}
Recent evidence that almost every local spheroid contains a super-massive black hole (SMBH) 
in its center whose mass is proportional to that of its host 
(e.g., Magorrian et al. 1998; Gebhardt et al. 2000; Ferrarese \& Merritt 2000) suggests that the growth 
of black holes and their host galaxies is mostly coeval. These observational results 
have found further support from hydro-dynamical simulations of galaxy formation, where active galactic nuclei (AGN) 
provide, mainly through winds and outflows, feedback between SMBHs and host spheroids 
(e.g., Cavaliere \& Vittorini 2000; Di Matteo, Springel \& Hernquist 2005; Menci et al. 2006, 2008 
and references therein). 
In this context, the most likely scenario is one in which most of the massive galaxies 
pass through a luminous starburst phase 
(with star-formation rates SFR of a~few~hundred~\msun~yr$^{-1}$) before the SMBH has built sufficient 
mass (10$^{8-9}$~\msun\ for a 10$^{12.4-13.4}$~\msun\ dark-matter halo; Granato et al. 2004) 
to shine as a luminous quasar (e.g., Silk \& Rees 1998; Fabian 1999; Hopkins et al. 2006). 
Initially the quasar would be obscured by the star formation that feeds it, 
but may eventually clear away gas and dust through various feedback mechanisms. 
Under this hypothesis, the fraction of obscured AGN can be interpreted as a measure of the timescale 
over which the nuclear feedback is at work (for a theoretical and observational investigation of this topic, 
see Menci et al. 2008; Fiore et al. 2009, and references therein). 
The SMBH emission will subsequently decline to leave a quiescent spheroidal galaxy. 
Sources caught during the initial luminous starburst phase may comprise the bulk of the sub-millimeter (sub-mm) galaxies 
discovered by \scuba\ over the last decade (e.g., Smail et al. 1997; Hughes et al. 1998), 
many of which being ultra-luminous infrared galaxies (ULIRGs; $L>10^{12}$~$L_\odot$) at $z\approx$~1--3 
(e.g., Blain et al. 2002); see also Sanders et al. (1988) for a similar scenario applied to ULIRGs and luminous quasars. 

Although the first investigations of AGN activity in sub-mm sources showed little overlap between the 
\xray\ and the sub-mm source populations (e.g., Fabian et al. 2000; Severgnini et al. 2000; 
Almaini et al. 2003; Waskett et al. 2003), subsequent more sensitive \xray\ observations in 
the \chandra\ Deep Field-North (CDF-N) 
were able to detect weak \xray\ counterparts of sub-mm sources (e.g., Alexander et al. 2003, 2005a, 2005b). 
These studies, coupled with the availability of spectroscopic information (e.g., Chapman et al. 2003, 2005), 
revealed that the AGN contribution to the sub-mm emission is of the order of a~few~per~cent, and that AGN 
in typical sub-mm galaxies are growing almost continuously, although the SMBHs in these galaxies are possibly 
several times less massive than seen in local massive galaxies (Alexander et al. 2008). 

Over the last few years, sub-mm evidence for the coeval growth of SMBHs 
and galaxy bulges has been reported by Page et al. (2001, 2004) and Stevens et al. (2004, 2005) 
from \scuba\ observations of \xray\ obscured AGN (spectroscopically classified as broad-line quasars) 
at $1<z<3$, which appear to have higher sub-mm output than a matched sample of unabsorbed quasars. 
The emerging observational picture is that a significant fraction of the long-wavelength 
radiation has to originate in a strong starburst, thus suggesting that these objects are 
simultaneously building up their central SMBH, being obscured by large amount of dust and gas, 
and stars, the estimated duration of the obscured growth phase being $\approx$~15~per~cent 
of the unobscured phase (Page et al. 2004; Stevens et al. 2005). If this picture were true, the unification model 
for AGN (Antonucci 1993), which explains the Type~1 (i.e., broad-line) vs. Type~2 
(i.e., narrow-line) AGN dichotomy as due to geometric effects, 
would necessarily need to be considered in a broader context, where AGN evolution and feedback mechanisms 
are important elements (e.g., Menci et al. 2008). 

% Coord [on the co-added R-band image produced by Marco M.]: 00:33:57.16 -12:00:38.26
To investigate the topics discussed above, we chose a sample of 16 luminous obscured quasars 
at $z$~$\approx$1--2 detected in the \hbox{2--10~keV} band in the HELLAS2XMM survey 
(see $\S$\ref{h2xmm_survey}), for which \spitzer\ follow-up observations have been performed 
(Pozzi et al. 2007; Pozzi et al., in preparation). 
Approximately half of this sample has a spectroscopic redshift and classification (Fiore et al. 2003; 
Maiolino et al. 2006; Cocchia et al. 2007); 
while IRAC and MIPS (24~\micron) data are available for the entire sample, 
\spitzer\ observations at 70~\micron\ and 160~\micron\ have been carried out only for eight sources 
with optical spectra. Finally, sub-mm observations have been performed for four of these sources only. 

Among the sample of 16 \xray\ selected sources, 
H2XMMJ~003357.2$-$120038 (RA=00 33 57.2; DEC=$-$12 00 38.3; hereafter \gd), 
a Type~2, radio quiet (S$_{1.4~\rm GHz}<0.13$~mJy at the 3$\sigma$ level) 
quasar at $z$=1.957, because of its broad-band coverage (up to 850~\micron) 
with good-quality photometric data, represents an ideal target to 
investigate the presence of both obscured AGN growth and star-forming activity at 
high redshift through a multi-band approach, and to estimate 
the bolometric energy output related to the accretion activity. 
In the following we will report on the multi-wavelength 
observational campaign conducted by our group over the last four years ($\S$2) and 
on the modeling of the spectral energy distribution (SED) of this source ($\S$3). 
The results will be discussed in $\S$4 and summarized in $\S$5. 

We note that \gd\ represents one of the few \xray\ selected obscured AGN at high-redshift 
with such multi-wavelength characterization [in particular, with sub-mm detection; 
see also Mainieri et al. 2005 for a $z$=3.660 AGN in the \chandra\ Deep Field-South (CDF-S)]. 

Hereafter we adopt \hbox{$H_{0}$=70~km~s$^{-1}$~Mpc$^{-1}$} in a $\Lambda$-cosmology 
with \hbox{$\Omega_{\rm M}$=0.3} and \hbox{$\Omega_{\Lambda}$=0.7} (Spergel et al. 2003). 
Magnitudes are reported in the Vega system.

\section{Multi-wavelength data reduction and analysis}
\label{multilambda}

\subsection{\gd\ in the context of the HELLAS2XMM survey}
\label{h2xmm_survey}
\gd\ was detected by the HELLAS2XMM survey\footnote{See the web page http://www.bo.astro.it/$\sim$hellas/sample.html.}, 
which comprises serendipitously 
detected sources in archival \xmm\ observations over (originally) 3~deg$^{2}$ (Baldi et al. 2002), 
with 2--10~keV fluxes in the $\approx$~6$\times10^{-15}$--4$\times10^{-13}$~\cgs\ range. 
The final source catalog of 232 \xray\ sources over 1.4~deg$^{2}$ 
(214 with optical counterpart down to R-band magnitude of $\approx$~25) is 
$\approx$~70~per~cent spectroscopically identified (Cocchia et al. 2007) and comprises both 
Type~1 and Type~2 AGN, emission-line galaxies, and early-type galaxies with indications of AGN activity from X-rays 
(XBONGs; see Comastri et al. 2002; Civano et al. 2007). 
In particular, albeit of low quality, the \xray\ spectrum of \gd\ was found consistent 
with being obscured by a rest-frame column density of 
7.3$^{+11.7}_{-5.5}\times10^{22}$~cm$^{-2}$ (Perola et al. 2004), once the photon index $\Gamma$ 
was fixed to a ``canonical'' AGN value of 1.9 (e.g., Piconcelli et al. 2005). 
The observed 2--10~keV flux of 2.3$\times10^{-14}$~\cgs, corresponding to an 
intrinsic rest-frame 2--10~keV luminosity ($\approx6.3\times10^{44}$~\lum), 
coupled with the presence of \xray\ absorption, 
classifies \gd\ as an \xray\ luminous obscured quasar. 
Finally, we note that the \xray\ luminosity is slightly higher ($\approx$~20~per~cent) 
than L$^{\star}$ (at $z=2$) from the La Franca et al. (2005) \xray\ luminosity function. 
% F[2-10 keV]=2.3+/-0.9E-14 (Perola) vs. 3.1E-14 (Fiore)

%%%%%%%%%%%%%%%%%%%%%%%%%%%%%%%%%%%%%%%%%%%%%%%%%%%%%%%%%%%%%%%%%%%%%%%%%%%%%%%%%%%%%%%%%%%%%%%%
%%%	FIGURE 1: smoothed optical spectrum [grism 13: 3685-9315, disp=2.77 A/pix, res=21.2 A]
%%%%%%%%%%%%%%%%%%
\begin{figure}
\includegraphics[angle=-90,width=0.50\textwidth]{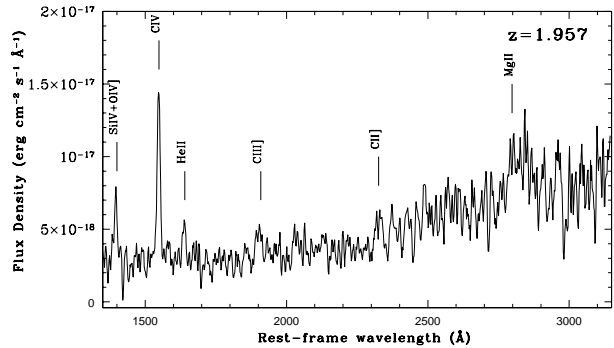}
\vglue-0.9cm
\caption{
Optical spectrum of \gd\ smoothed with a boxcar of 16\AA; the main emission features are labeled 
(ESO, EFOSC2, 30-minute exposure).}
\label{optical_spectrum}
\end{figure}
%%%%%%%%%%%%%%%%%%
%%%	END of FIG. 1
%%%%%%%%%%%%%%%%%%%%%%%%%%%%%%%%%%%%%%%%%%%%%%%%%%%%%%%%%%%%%%%%%%%%%%%%%%%%%%%%%%%%%%%%%%%%%%%%
%
%%%%%%%%%%%%%%%%%%%%%%%%%%%%%%%%%%%%%%%%%%%%%%%%%%%%%%%%%%%%%%%%%%%%%%%%%%%%%%%%%%%%%%%%%%%%%%%%
%%%	FIGURE 2: Spitzer images [4xIRAC + 3xMIPS]
%%%%%%%%%%%%%%%%%%
\begin{figure*}
\includegraphics[angle=0,width=0.9\textwidth]{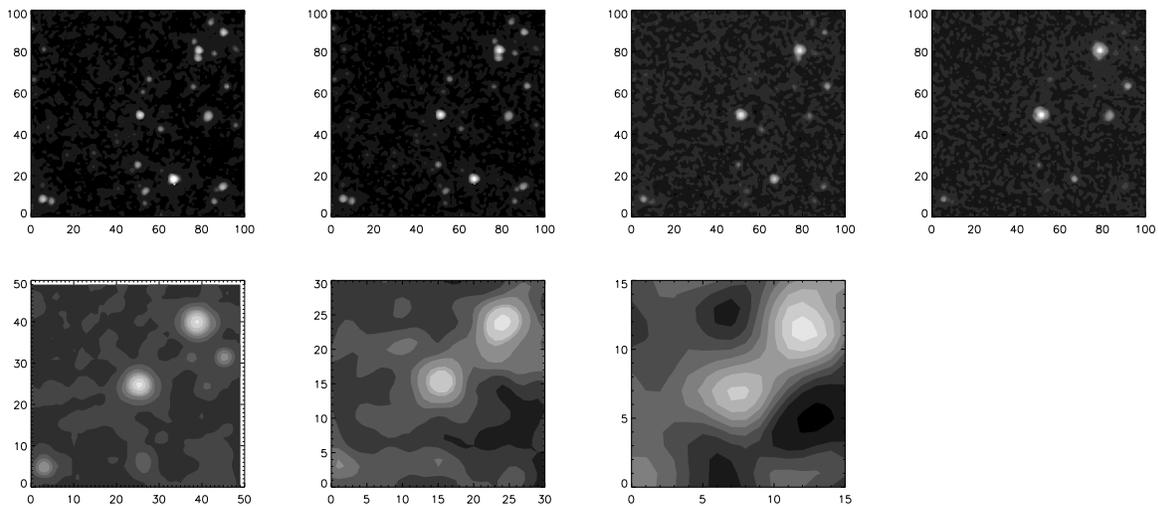}
\caption{
Cut-outs of the \spitzer\ images centered on \gd: {\it (top)} IRAC at 
3.6~\micron, 4.5~\micron, 5.8~\micron, and 8~\micron; 
{\it (bottom)} MIPS at 24~\micron, 70~\micron, and 160~\micron. 
The size of each box is 2$\times$2~arcmin$^{2}$. 
North is up, and East to the left, rotated counterclockwise by an angle 
of 25.8 degrees (optimal angle adopted by {\sc mopex} in the process of merging images). 
The labeled numbers in each box are pixel units.}
\label{spitzer_images}
\end{figure*}
%%%%%%%%%%%%%%%%%%
%%%	END of FIG. 2
%%%%%%%%%%%%%%%%%%%%%%%%%%%%%%%%%%%%%%%%%%%%%%%%%%%%%%%%%%%%%%%%%%%%%%%%%%%%%%%%%%

\subsection{Optical spectroscopic classification}
\label{optical_data}
% Spect-ID observation: 21/08/01; photometry: 18/08/01
Optical (R-band) photometry and long-slit spectroscopy (grism \#13, covering a range of 
$\approx$~3700-9300~\AA) of \gd\ were obtained in 2001 
at the ESO 3.6m (EFOSC2 instrument), for a total exposure time of 10~minutes and 30~minutes, respectively; 
for details, see Fiore et al. (2003). 
The object is relatively bright in the optical (R=21.8); all of the clearly identified permitted emission 
lines (\civ\ and \heii) have a full width at half maximum (FWHM) $<$1500~km~s$^{-1}$. 
In the noisy, red part of the spectrum (see Fig.~\ref{optical_spectrum}) there are 
some indications of a broad \mgii\ emission line, which is not 
centered at the expected rest-frame wavelength of 2798\AA. 
The emission-line shift between the broad \mgii\ and the narrower \civ\ 
would be unusually large ($\approx$~9000 km~s$^{-1}$ of redshift of the \mgii\ with respect to the 
emission features in the blue part of the spectrum). This shift is three times as large as the 
maximum shift found by Richards et al. (2002) in a large sample of SDSS quasars. 
%
%However, this apparently broad feature is not 
%centered at the expected rest-frame wavelength of 2798\AA, suggesting either some spurious contribution 
%from bad pixels or residuals from background subtraction (or both).
%
The current data suggest that either some spurious contribution from bad pixels redward of the \mgii\ line or residuals 
from background subtraction (or both) mimic a spurious broad emission line. 
Furthermore, also the redness of the continuum, the equivalent widths of the lines and their ratios, 
which are different from the values obtained from the composite Type~1 quasar spectrum obtained from 
the SDSS (Vanden Berk et al. 2001), are all suggestive of a Type~2 AGN. 
We also note that \gd\ has different properties with respect to those of the \xray\ obscured quasars 
at high redshifts found by Page et al. (2001, 2004; see also Stevens et al. 2005), 
which are characterized by broad-line optical spectra.  
Therefore, both the \xray\ and optical properties place \gd\ in the long sought-after 
class of ``genuine'' Type~2 quasars (see also Comastri, Vignali \& Brusa 2002).

%%%%%%%%%%%%%%%%%%%%%%%%
\input{vignali.tab1.tex}
%%%%%%%%%%%%%%%%%%%%%%%%

\subsection{Spitzer data}
\label{spitzer_data}
\gd\ was observed by \spitzer\ in 2006 with both IRAC and MIPS instruments 
in photometry mode, for a total integration time of 480~s (IRAC), 1400~s (MIPS at 24~\micron), 
3000~s (MIPS at 70\micron), and 600~s (MIPS 160~\micron). 
For the IRAC bands, we used the final post-calibrated data (PBCD) produced by the \spitzer\ Science Center (SSC) pipeline. 
%(Version S14.0.0). 
For the typically more ``problematic'' MIPS bands, we started the analysis from the basic calibrated data 
(BCD) and we constructed our own mosaics using the SSC {\sc mopex} software (Makovoz \& Marleau 2005). 
At 24~\micron, the input BCD for {\sc mopex} were previously corrected using ``ad-hoc'' procedures 
to remove the residual flat fielding, dependent on the scan mirror position 
(see also Fadda et al. 2006; Pozzi et al. 2007). 
At 70 and 160~\micron, we used the median high-pass filtered BCD (FBCD) 
to produce the mosaic map, as suggested for faint point-like sources (see the MIPS Data Handbook, Version 3.3.0). 

The flux densities in the IRAC and MIPS bands were measured on the signal maps using aperture photometry. 
The chosen aperture radius for the IRAC bands was 2.45\arcsec\ and the
adopted factors for the aperture corrections are 1.21, 1.23, 1.38 and 1.58 (following the IRAC Data
Handbook, Version 3.0) at 3.6~\micron, 4.5~\micron, 5.8~\micron, and 8~\micron, respectively. 

The chosen aperture radius for the MIPS bands was 7\arcsec\ 
at 24~\micron\ and 16\arcsec\ at 70~\micron\ and 
160~\micron\footnote{The PSF FWHM in the MIPS bands is $\approx$~6\arcsec\ (24~\micron), 18\arcsec\ (70~\micron), 
and 40\arcsec\ (160~\micron); see MIPS Data Handbook.}; 
the aperture flux densities were corrected adopting a factor of 1.61, 2.07 and 4.1 in the three MIPS bands, respectively 
(see the MIPS Data Handbook). At 70~\micron\ and 160~\micron, these 
factors are well suited for sources with dust temperature of T$\approx$50--60~K. 
At these wavelengths, we were forced to use a small extraction radius (in particular at
160~\micron, where the radius is comparable to half of the PSF FWHM), since at a distance of 
$\approx$~50$^{\prime\prime}$ from our target there is a source (undetected in the \xray\ band) 
of comparable brightness in the IRAC bands but $\approx$~35 and 20~per~cent brighter 
at 70~\micron\ and 160~\micron, respectively (Fig.~\ref{spitzer_images}).  
The companion source would contaminate the target emission if larger apertures were adopted. 
Using an aperture diameter smaller than the PSF FWHM (as in the case
of 160~\micron\ data) means to rely on the PSF core, where flux non-linearities 
could play a significant effect due to the limited sampling of the PSF (see Gordon et al. 2007). 
On the other hand, we have tested how a PSF fitting would not provide reliable results, 
probably due to the small field-of-view of the mosaics 
($\approx$~2--3~arcmin; see also Stansberry et al. 2007, where the latest 160~\micron\ flux density 
absolute calibration is discussed). 

Table~\ref{multi_photo} reports the target flux densities provided by \spitzer. 
To compute the uncertainties, we added in quadrature the noise map and the systematic uncertainty, 
assumed conservatively to be 10~per cent in the IRAC and MIPS 24~\micron\ bands, 
and 15~per~cent at 70~\micron\ and 160~\micron\ (see IRAC and MIPS Data Handbooks). 
While the systematic uncertainties dominate in the IRAC and MIPS 24~\micron\ bands, at longer wavelengths also 
the noise map is an important factor of uncertainty, approaching the level of the systematic uncertainty at 160~\micron. 

We note that, given our flux density measurements in the IRAC bands, 
\gd\ would have been recognized as an AGN in both the 
Lacy et al. (2004) and Stern et al. (2005) diagnostic diagrams. 
Also the companion source would lie in the AGN ``locus'' in these diagrams. 
Given the source properties in the mid-IR and far-IR, it appears plausible that 
\gd\ and its companion are both characterized by AGN and starburst emission. 
The \xray\ non-detection of the companion source may indicate either weak \xray\ emission or obscuration, 
although the limited exposure of the \xmm\ 
observation ($\approx$~21.6~ks and 13.8~ks in the MOS1 and MOS2, respectively) 
prevents from further investigation. 

Using the 24~\micron\ galaxy counts in the SWIRE fields from Shupe et al. (2008), we computed the integral 
number density of 24~\micron\ sources with flux density above $\approx$~5.3~mJy (the flux densities are 
5.33~mJy and 5.20~mJy for \gd\ and the companion, respectively), which is $\approx$~25~deg$^{-2}$ 
(a similar number density is found in Papovich et al. 2004); 
this number translates into 0.015 expected sources in a circular region with radius of 50\arcsec\ 
(the distance between the two sources). 
The Poisson probability of obtaining one or more sources when 0.015 are expected is relatively low 
($\approx1.49\times10^{-2}$); if placed at the same redshift, they could represent the density peaks in 
an over-dense region, as found in the fields around some high-redshift radio galaxies with associated 
powerful sub-mm emission (e.g., Stevens et al. 2003; Smail et al. 2003). 

\subsection{SCUBA data}
\label{scuba_data}
% Nights: 06 Oct 2004 - 11 Nov. 2004 for \gd
% Nights: 04 Oct, 2004 - 06 Oct, 2004 (first set of \gd\ obsn) - 11 Nov, 2004 (II set of \gd\ obsn) - 19 Dec, 2004
% Calibrators: crl2688 - Uranus (for \gd)
The \scuba\ observations of \gd\ were split in two nights over October--November 2004 and carried out 
in photometry mode, in which the source is placed on the central bolometer of the array, 
with the wide 850:450 filter set. 
The sky opacity was monitored on a quasi-continuous basis with the JCMT water vapour radiometer and cross-checked 
with skydips several times per night. Uranus was chosen as flux-density calibrator; 
the estimated calibration uncertainties on the most sensitive 850~\micron\ data are $\approx$~10\%. 
Data were reduced with the Starlink {\sc surf} package in a standard manner (e.g., Holland 
et al. 1999). 
We reduced data from the two nights separately and then calculated a weighted mean flux density and error. 
The source was detected at the 4.2$\sigma$ level at 850~\micron, while at 450~\micron\ only a loose 
upper limit was obtained (reported at the 3$\sigma$ level in Table~\ref{multi_photo}). 
Although the \spitzer\ 160~\micron\ data reveal the presence of a bright infrared ``companion'' source, 
its distance from \gd\ is such that it should not give any significant flux contribution 
to our target at 850~\micron\ given the \scuba\ beam size ($\approx$~15\arcsec) at this wavelength.

\section{Data modeling}
\label{model_description_modeling}
The optical-to-sub-mm data of \gd\ have been modeled using the code developed by Fritz et al. (2006, hereafter F06). 
This code assumes a continuous dust distribution (around the central source) consisting of silicate and 
graphite grains, confined in a toroidal shape (Pier \& Krolik 1992), 
as opposed to the ``clumpy'' tori models mostly developed 
over recent years (e.g., Nenkova, Ivezi{\'c} \& Elitzur 2002; H{\"o}nig, Beckert \& Ohnaka 2006; 
Nenkova et al. 2008a,b; Schartmann et al. 2008). 
Notwithstanding this probably simplified underlying modeling, the fundamental goal of the present 
work consists in estimating the physical parameters related to the AGN (torus) emission, peaking in the mid-IR domain. 
We adopted the multi-component fit approach (e.g., Hatziminaoglou et al. 2008); in the following, 
we resume its main characteristics. 

We use three distinct components that, in general, dominate different parts of a galaxy SED: 
\begin{enumerate}
\item stellar emission (UV--near-IR); 
\item emission due to hot dust heated by AGN emission (mid-IR); 
\item emission due to warm/cold dust heated by star formation (starburst component, mid- and far-IR). 
\end{enumerate}

(i) 
To model the stellar component, we use a set of Simple Stellar Population (SSP) spectra of solar 
metallicity and ages ranging from $\approx 1$~Myr to 1.7~Gyr; 
the latter is the time elapsed between $z$=4 (the redshift assumed for the stars to form) and $z=1.957$ 
in the adopted cosmology. 
A Salpeter (1955) initial mass function (IMF) has been assumed. 
These are weighted by a Schmidt-like law of star formation:
\begin{equation}\label{eqn:schmidt}
SFR(t)=\frac{T_G-t}{T_G} \exp\left(\frac{T_G-t}{\tau\cdot T_G}\right)
\end{equation}
where $T_G$ is the age of the oldest stellar population and $\tau$ sets the duration of the initial burst 
(normalized to $T_G$). 
%A Schmidt-like law (i.e., with an exponential decline) has been assumed for the star formation, 
%and the Galactic extinction law (R$_V=3.1$; Cardelli, Clayton \& Mathis 1989) has been adopted for the stars of all ages. 
A common value of extinction is applied to stars of all ages, and the extinction law of the 
Galaxy has been adopted (R$_V=3.1$; Cardelli, Clayton \& Mathis 1989). 

(ii) Type~2 AGN emission, dominating in the mid-IR domain 
(from a rest-frame wavelength of $\approx$~2~\micron\ up to $\approx$~30--50~\micron) 
and produced in a dusty region surrounding the SMBH, is modeled using an extended version 
of the model grid described by F06. 

(iii) The emission from warm/cold dust, peaking in the far-IR (at $\approx$~100~\micron\ rest frame) and being characterized by 
features in the mid-IR due to PAH molecules, has been accounted for by using semi-empirical models of 
known and well studied starbursts (Arp~220, M~82, M~83, NGC~1482, NGC~4102, NGC~5253 and NGC~7714). 

The fit to the observed photometry is carried out at first on the UV-optical-near-IR region 
(over the $\approx~0.3-5$~\micron\ rest-frame interval), 
by means of a pure stellar component. 
Then all the starburst templates are normalized to match the far-IR data (longward of $\approx$~25~\micron\ rest frame). 
At this point, the best AGN model is searched, among all of those available from the aforementioned model grid, 
by varying the torus parameters (e.g., effective angle, viewing angle, dust distribution, etc.). 
Finally, the normalizations of the three components are iteratively changed until a good global solution is achieved; 
at this stage, also the possibility that one of these three components may not be necessary is taken into account. 
The goodness of the fit is evaluated by means of a ``merit function'' which may be intended as a modified reduced $\chi^2$, where the 
number of degrees of freedom is replaced by the number of photometric points used in the SED fitting, following the equation 
\begin{equation}\label{eqn:chi2}
MF=\sum_{i=1}^{N_O}\left(\frac{M_i-O_i}{\sigma_i}\right)^2/N_O
\end{equation}
where $N_O$ is the number of observed datapoints, $M_i$ and $O_i$ are the model and observed flux densities, respectively, 
of the $i-th$ photometric band, and $\sigma_i$ is the corresponding observed error. 
Since the observed SED at study includes only photometric data and, furthermore, is not fully sampled, 
the problem of finding a best-fit model is intrinsically prone to a certain degree of degeneracy, i.e. 
to the non-uniqueness of a solution. 

As {\em a posteriori} check, we estimated what \aox\ 
(defined as the slope of a hypothetical power-law connecting the rest-frame 2500~\AA\ to 2~keV 
($\alpha_{\rm ox}=\frac{\log(f_{\rm 2~keV}/f_{2500~\mbox{\rm \scriptsize\AA}})}{\log(\nu_{\rm 2~keV}/\nu_{2500~\mbox{\rm \scriptsize\AA}})}$) 
is needed to reproduce the source rest-frame, de-absorbed 2--10~keV luminosity. 
This will allow us to provide a comparison with recent literature results 
based on optically selected broad-line quasars (e.g., Vignali, Brandt \& Schneider 2003; Steffen et al. 2006).

\section{Results from the SED fitting and discussion} 
\label{results_sed_fitting}
The output of the SED fitting procedure described above consists of a series of models with different $MF$, each having 
its own set of parameters describing the stellar, AGN and starburst components. 
In the following, we will describe the best-fitting model  
parameters related to the SED of \gd. 

%%%%%%%%%%%%%%%%%%%%%%%%%%%%%%%%%%%%%%%%%%%%%%%%%%%%%%%%%%%%%%%%%%%%%%%%%%%%%%%%%%%%%%%%%%%%%%%%
%%%	FIGURE 3: SCUBA source' SED
%%%%%%%%%%%%%%%%%%
\begin{figure}
\includegraphics[angle=-90,width=0.5\textwidth]{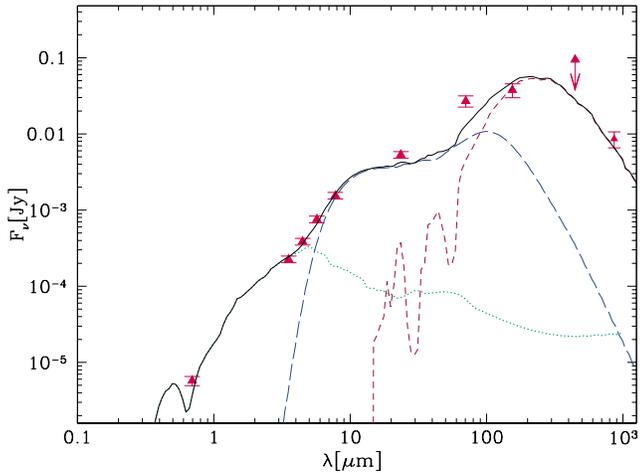}
\caption{
Observed-frame SED of \gd\ from the R-band to the \scuba\ 850~\micron\ data points. 
The upper limit refers to the source non-detection at 450~\micron\ (\scuba\ data). 
The ``global'' SED (solid line), reproduced as the summed contribution of 
a stellar component (dotted line), a torus component (long-dashed line), and a starburst component (short-dashed line), 
represents the best-fitting according to F06 modeling.}
\label{sed_lx_aox16}
\end{figure}
%%%%%%%%%%%%%%%%%%
%%%	END of FIG. 3
%%%%%%%%%%%%%%%%%%%%%%%%%%%%%%%%%%%%%%%%%%%%%%%%%%%%%%%%%%%%%%%%%%%%%%%%%%%%%%%%%%
%
The best fitting has a $MF$ of $\approx$~2.2; most of the discrepancy is due to the modeling over the wavelength 
range 24--70~\micron\ (see Fig.~\ref{sed_lx_aox16}), where the model SED is a factor $\approx$~2 below the photometric points. 
However, we note that apparently unacceptable $MF$s represent a common problem of most of the SED fittings 
(see, e.g., Fig.~2 of Hatziminaoglou et al. 2008), due also to the combination of 
photometric measurements with largely different accuracies (see, for a comprehensive description of this issue, 
Gruppioni et al. 2008). 

The AGN component in the best-fitting ``global'' SED model (solid line in Fig.~\ref{sed_lx_aox16}) consists of reprocessed 
emission by an almost edge-on torus ($\approx$~10~degrees with respect to the equatorial plane) -- as expected from the optical 
and \xray\ classifications of \gd\ as a Type~2 quasar -- 
with a full covering angle\footnote{We note that in F06 this angle was referred to as full opening angle.} 
of $\approx$~140~degrees (corresponding to a covering factor of $\approx$~90~per~cent) 
and an optical depth (at 9.7~\micron) 
of $\tau(9.7)=1.0$, corresponding to a column density of $\approx$~9$\times10^{22}$~cm$^{-2}$ for a Galactic dust-to-gas ratio. 
This value is similar to the one derived (though with large uncertainties) from the \xray\ spectrum. 
Given the source \xray\ luminosity, the corresponding optical-to-X-ray spectral slope would be \aox=$-1.55\div-1.60$. 

The AGN contribution to the rest-frame 1--1000~\micron\ emission is 54~per~cent; the remaining emission is mostly ascribed to 
a prominent starburst emission, best reproduced by an Arp~220 template (Berta et al. 2003; short-dashed line in Fig.~\ref{sed_lx_aox16}), 
made by combining {\sc grasil} (Silva et al. 1998) models and the 5--20~\micron\ ISOCAM-CVF spectrum of the source (Charmandaris et al. 1999). 
Despite this template under-estimates the observed flux density at 70~\micron\ (close to the peak of the torus emission), 
we note that no other starburst template is able to reproduce the far-IR and sub-mm emission of \gd\ better than Arp~220. 

From the infrared luminosity -- once the AGN contribution has been removed -- and adopting the Kennicutt (1998) relation, we 
obtain a SFR of $\approx$~1500~\sfr. 
For comparison and further check, the recent SFR was also computed using Eq.~(1), 
i.e., integrating the law of star formation through fitting the rest-frame optical/near-IR datapoints with the stellar component 
(dotted line in Fig.~\ref{sed_lx_aox16}), characterized by \ebv$\approx$0.49. 
The derived SFR is consistent with that obtained from the Kennicutt (1998) relation. 

The SFR for \gd\ appears to be consistent with the values derived by Stevens et al. (2005) for a sample of 
broad-line, \xray\ obscured quasars detected with \scuba\ over the redshift range \hbox{$z$$\approx$1--3} 
and with the sub-mm detected quasars at $z\approx$~2 from Coppin et al. (2008). 
For comparison, we note that the SFR derived by Mainieri et al. (2005) for an \xray\ selected Type~2 quasar at $z$=3.66 
is a factor $\approx$~2.2--2.5 lower. 
At zero-th order, there are significant indications that \gd\ is experiencing 
a phase of strong star formation and AGN activity, similarly to other quasars at $z\approx2$. 

No other SED fitting solution using the three-component modeling provides similar results in terms of MF and 
overall representation of the \xray\ properties of \gd. 
While in all of the SED fittings the full covering angle of the torus is 140~degrees, 
solutions with $\tau(9.7)<1.0$ (corresponding to higher inclination angles 
for the line of sight with respect to the equatorial plane of the torus) 
would significantly under-estimate the observed column density. 
Similarly, a solution requiring a flatter \aox\ ($-$1.30, close to the average \aox\ found in \xray\ selected 
broad-line quasars in the XMM-COSMOS survey; Lusso et al., in preparation) would imply a worse MF ($\approx$~4.4), 
mostly due to its inability in reproducing the source emission over the wavelength range \hbox{5--8~\micron} 
(where the stellar and AGN components are approximately equal); in this case, the parameters derived for the AGN and 
starburst emission are similar to those presented above. 
We also note that solutions with only the AGN plus starburst emission can be discarded on the basis of 
the low optical depth ($\tau(9.7)$=0.09, i.e., a factor of $\approx$~10 lower column density than 
actually measured); 
in this case, a steeper \aox, close to $\approx$~$-$1.8, would be required to reproduce the observed SED, 
and a factor $\approx$~4 higher bolometric luminosity than that obtained with the best-fitting solution 
(see below) would be implied. 

Despite the fact that $\tau(9.7)$ along the line of sight is equal to 1, 
the 9.7~\micron\ silicate feature is seen in emission in the best-fitting model, while one would expect it in absorption. 
This is due to the combination of two effects: first, having a lower sublimation temperature with respect to 
graphite, the silicate grains can only survive at larger distances from the inner radius of the torus. 
Thus the optical depth, calculated from this ``hot-silicate'' zone towards the observer, 
can be, in some cases, lower than unity (because not integrated throughout the entire thickness of the torus), 
thus making possible the observation of this feature in emission. 
Second, in models where the density is decreasing at increasing height over the equatorial plane, 
the emission feature comes from the lower density zones, where the hot silicates are living in an optically thinner environment.
%
%%%%%%%%%%%%%%%%%%%%%%%%%%%%%%%%%%%%%%%%%%%%%%%%%%%%%%%%%%%%%%%%%%%%%%%%%%%%%%%%%%%%%%%%%%%%%%%%
%%%	FIGURE 4: SCUBA source' SED
%%%%%%%%%%%%%%%%%%
%\begin{figure}
%\includegraphics[angle=-90,width=0.5\textwidth]{fit_3comp_Lx_aox13_chi4.4_ext.eps}
%\caption{
%Observed-frame SED of \gd, similarly to Fig.~\ref{sed_lx_aox16}; in this case, \aox=$-$1.3 has been assumed. 
%}
%\label{sed_lx_aox13}
%\end{figure}
%%%%%%%%%%%%%%%%%%
%%%	END of FIG. 4
%%%%%%%%%%%%%%%%%%%%%%%%%%%%%%%%%%%%%%%%%%%%%%%%%%%%%%%%%%%%%%%%%%%%%%%%%%%%%%%%%%

According to recent results, a limited fraction of optically luminous quasars at $z\approx2$ are detected in the 
sub-mm and millimeter (mm) wavelengths ($\approx$~25~per cent; Omont et al. 2003), 
suggesting that the quasar and sub-mm phases do not overlap significantly, probably because of the different 
lifetimes of the two phases. 
Within this scenario, a quasar detected in the sub-mm/mm is a good candidate for being caught in the so-called 
``transition phase'' from a growing, sub-mm galaxy to an unobscured quasar. In such a phase 
(whose duration is estimated of the order of $\approx$~15~per~cent of the unobscured phase; Stevens et al. 2005), 
powerful obscured emission from the active nucleus, coupled with a prominent starburst component, 
is expected and actually observed (e.g., Page et al. 2001, 2004; Stevens et al. 2005; see also Aravena et al. 2008). 

We estimate an AGN bolometric luminosity, integrated over the optical, UV and \xray\ band (extrapolated up to 
500~keV assuming a power law with photon index $\Gamma$=1.8 and a cut-off at 300~keV), of 
$\approx$~4.3$\times10^{46}$~\lum. 

Since the black hole mass of \gd\ cannot be derived using ``standard'' methods (e.g., reverberation, scaling relationship, etc.), 
we decided to estimate it using the M$_{\rm BH}$--M$_{\rm bulge}$ relation derived locally by Marconi \& Hunt (2003) from a sample of 
$\approx$~30 galaxies with a ``secure'' black hole mass measurement (hence, no evolution of the relation over cosmic time is 
assumed). 
% No evolution in Mbh/Mbulge relation; Lk actually evolves...
The stellar mass, 
obtained directly from the best-fitting SED modeling, is $\approx1.5\times10^{12}$~\msun, 
close to the upper boundary for the stellar masses reported by Dye et al. (2008; see their Fig.~9) 
for a sample of \scuba\ galaxies at $z\approx2$. 
The assumption of a Chabrier (2003) IMF, which has been suggested to be more representative of the IMF at high redshifts 
(e.g., Renzini 2006), would imply a factor $\approx$~1.7 lower stellar masses, without significant changes in the overall 
SED fitting. 
%%%%%%%%%%%%%%%%%%%%%%%%%%%%%%%%%%%%%%%%%%%%%
% Eddington luminosity
% Ledd=1.23*E38 M/Msun --> 2.34E47 erg/s
% L(1.24-300 keV)=3.6E45 erg/s
% L[accr, model (a)]=3.98E46 
%              + Lx =4.34E46 erg/s
% L[accr, model (b)]=6.7E45 erg/s
%              + Lx =1.03E46 erg/s
% --> Eddington ratio=0.19-0.04
% Model with AGN+STB only: Lacc=1.75E47 erg/s
%                          +Lx =1.79E47 erg/s
% --> Eddington ratio=0.76
%%%%%%%%%%%%%%%%%%%%%%%%%%%%%%%%%%%%%%%%%%%%%
The estimated black hole mass is $\approx1.9\times10^{9}$~\msun, implying an Eddington ratio of $\approx$~0.19, 
higher than the values recently found for a sample of HELLAS2XMM selected Type~2 quasars at high redshift 
($\approx$~0.05 on average; Pozzi et al. 2007; Pozzi et al., in preparation; for further 
details on the source selection, see also $\S$\ref{introduction}). 

If we assume that the star-formation episode lasts, at the rate estimated in this paper, for a time interval $t$, while the black hole
grows at the current rate for a period $t_{\rm qso}=\eta\times t$ (where $\eta$ depends on the presence of available fueling for 
accretion, hence on the history of mergers and dynamical instabilities of \gd), we derive that the system would move $\approx$~0.3~dex 
above (below) the Marconi \& Hunt (2003) relation over a timescale of $\approx6\times10^{8}$~yr ($\approx4\times10^{9}$~yr) if $\eta$=1.0 (0.1) 
is assumed. In other words, the persistence of the system along the relation depends on the relative duration of the quasar and starburst; 
in the case of $\eta$=1.0, that represents the situation in which their phases last for a comparable amount of time 
(at the rates estimated in our work), it would take less than 1~Gyr to move the system away from the relation. 

% ----------------------------
% Lamastra et al. (2006) model
% ----------------------------
An alternative, and admittedly speculative, possibility is that the cold gas 
and dust reservoir, which sustain the star formation at the estimated rate of $\approx$~1500~\sfr, 
is also responsible for the obscuration of the nuclear source. 
The observed \xray\ absorption can be easily reproduced for a relatively large range of the 
Lamastra, Perola \& Matt (2006) molecular disk parameters, provided that the surface density exceeds 
$\approx$~8~\msun\ pc$^{-2}$. We note that the molecular gas has to be close to the nuclear region, 
otherwise the narrow-line region would be hardly visible. 

% -----------------------------------------------------
% 2-component (agn+stb) modeling of the SED [see above]
% -----------------------------------------------------
%Finally, we also checked whether different solutions for the overall SED could be obtained. 
%As an alternative possibility to the three-component SED fitting, we also ran the F06 code with only the AGN plus starburst emission. 
%In this case, all the optical, near-IR and mid-IR emission of \gd\ is reproduced by the AGN, without any contribution from stars. 
%Besides the physical reliability of such model, 
%the best-fitting SED ($MF\approx3.1$) would require a large value for \aox\ ($\approx-1.9$), 
%much higher (in absolute terms) than actually found from previous fittings and in \xray\ selected AGN, and would hardly 
%reproduce the observed \xray\ absorption, with $\tau(9.7)\simlt0.1$. In this SED modeling, the torus, characterized by a 
%full covering angle similar to the previous models ($\approx$~145~degrees), is seen at a higher inclination 
%angle with respect to the line of sight ($\approx$~50~degrees). The corresponding bolometric luminosity would be a factor $\approx$~4 
%higher than in {\sf model (b)}, implying an Eddington ratio of $\approx$~0.76. 
%%%%%%%%%

\section{Summary of the results}
We have presented the optical, infrared and sub-mm observations of a luminous, \xray\ obscured Type~2 quasar at $z$=1.957 
detected in the 2--10~keV band within the HELLAS2XMM survey.  
By fitting the photometric points with a three-component fit approach 
(stellar, AGN and starburst emission; see F06 for details), we were able to estimate some fundamental parameters related 
to the physical and geometrical properties of the torus and starburst components. 
\begin{description}
\item[$\bullet$]
The AGN component, related to emission reprocessed by dust in the torus, is dominant in the mid-IR, 
peaking at $\approx$~20--30~\micron\ (rest frame). 
The obscuring matter is likely distributed in an almost edge-on structure ($\approx$~10~degrees with respect to the equatorial plane), 
with a full covering angle of $\approx$~140~degrees. 
The derived optical depth [$\tau(9.7)=1.0$], corresponding to a column density of $\approx$~9$\times10^{22}$~cm$^{-2}$, 
is in agreement with the absorption derived by the \xray\ spectral fitting. 

\item[$\bullet$]
The AGN contribution to the rest-frame 1--1000~\micron\ emission is of $\approx$~54~per~cent, the remaining being ascribed to 
a starburst emission (best reproduced by an Arp~220 template) peaking at longer wavelengths with respect to the AGN emission. 
The derived SFR, $\approx$~1500~\sfr, places \gd\ as a breeding ground for considerable star formation at $z\approx2$. 

\item[$\bullet$]
%The picture emerging from the results of the SED fitting is that of an object caught in the coeval AGN and star-formation activity, when the 
%AGN is powerful but still obscured by the matter which has sustained its primordial growth. The AGN activity, however, is not powerful enough to 
%remove all the gas reservoir, thus depriving the star formation
The picture emerging from the results of the SED fitting is that of an object caught in a phase where intense, dust-enshrouded 
star formation is coeval to obscured AGN growth, which however does not prevent the active nucleus from revealing its presence 
in the optical and \xray\ bands.
If current evolutionary scenarios (e.g., Hopkins et al. 2006; Menci et al. 2008) were correct, 
we could expect that the black hole in \gd\ 
is in the process of becoming a ``classic'', luminous and unobscured quasar, capable of quenching the star formation. 
There are indications that such phase is shorter for the most luminous AGN (e.g., Adelberger \& Steidel 2005; Menci et al. 2008). 
\gd\ seems to have properties in common with the sub-mm detected and \xray\ obscured quasars at $z$$\approx$1--3 found recently in other surveys 
(Page et al. 2004; Stevens et al. 2005). 
%for these objects, the \xray\ obscuration has been ascribed to a dense ionised wind driven by the quasar (Page et al. 2006). 
\end{description}

\section*{Acknowledgments}
CV and AC thank for partial support the Italian Space Agency 
(contracts ASI--INAF I/023/05/0 and ASI I/088/06/0) and PRIN--MIUR (grant 2006-02-5203); 
CV, FP, AC, CG, FF and RM acknowledge partial support by ASI through contract ASI-INAF I/016/07/0. 
CV is very grateful to H. Butner, J. Cox, I. Coulson, D. Delorm, J. Kemp and B. Weferling for 
kind assistance in planning the \scuba\ observations and during the visit to JCMT facilities, 
M. Bolzonella for providing extinction curves, 
E. Diolaiti for help with source deblending techniques, D. Fadda for suggestions about MIPS data cleaning 
and reduction, P. Severgnini for help in sub-mm data reduction and fruitful discussions, A. Caccianiga for help with 
{\sc iraf} tools, A. Cimatti, R. Gilli, A. Lamastra and G.~C. Perola for helpful suggestions. 
The authors thank also F. Bauer for English editing, 
and the referee for his/her useful comments and suggestions. 
This work has benefited from research funding from the European Community's Sixth Framework Programme.

\end{document}

%% file: vignali.tab1.tex
%%%%%%%%%%%%%%%%%%%%%%%%%%%%%%%%%%%%%%%%%%%%%%%%%%%%%%%%%%%%%%%%%%%%%%%%%%
%%%	Table 1: GD 158_19: multi-band photometry
%%%%%%%%%%%%%%%%
\begin{center}
\begin{table}
%\centering
\begin{minipage}{0.5\textwidth}
\caption{\gd: optical--sub-mm photometry.}
\label{multi_photo}
\begin{tabular}{ccc}
\hline
Observed & Instr. & Flux Density \\
Band     &        & (\mjy)       \\
\hline
% R=21.80+/-0.15 --> 5.60+0.83/-0.72
R	     & ESO/3.6m      &   5.6$\pm{0.8}$   \\
3.6~\micron  & \spitzer/IRAC &   226$\pm{23}$  \\
4.5~\micron  & \spitzer/IRAC &   387$\pm{39}$  \\
5.8~\micron  & \spitzer/IRAC &   758$\pm{76}$  \\
8~\micron    & \spitzer/IRAC &  1547$\pm{155}$ \\
24~\micron   & \spitzer/MIPS &  5326$\pm{534}$ \\
70~\micron   & \spitzer/MIPS & 27000$\pm{4500}$    \\
160~\micron  & \spitzer/MIPS & 37700$\pm{7800}$    \\
450~\micron  & \scuba        & $<94300$            \\
850~\micron  & \scuba        &  8610$\pm{2060}$    \\
\hline
\end{tabular}
\end{minipage}
%\hglue-3.0cm
\begin{minipage}{5.7cm}
The upper limit to \scuba\ 450~\micron\ flux density is reported at 
the 3$\sigma$ confidence level. 
\end{minipage}
\end{table}
\end{center}
%%%%%%%%%%%%%%%%
%%%     End of Table 1
%%%%%%%%%%%%%%%%%%%%%%%%%%%%%%%%%%%%%%%%%%%%%%%%%%%%%%%%%%%%%%%%%%%%%%%%%%

%% file: vignali.bbl
\begin{thebibliography}{99}
%%%%%%%%%%%%
\bibitem[]{}
Adelberger K.~L., Steidel C.~C., 2005, \apj, 630, 50 
%
\bibitem[]{}
Alexander D.~M. et al., 2003, \aj, 125, 383 
%
\bibitem[]{}
Alexander D.~M., Bauer F.~E., Chapman S.~C., Smail I., Blain A.~W., Brandt W.~N., Ivison R.~J., 
2005a, \apj, 632, 736 
%
\bibitem[]{}
Alexander D.~M., Smail I., Bauer F.~E., Chapman S.~C., Blain A.~W., Brandt W.~N., Ivison R.~J., 
2005b, \nat, 434, 738 
%
\bibitem[]{}
Alexander D.~M. et al., 2008, \aj, 135, 1968 
%
\bibitem[]{}
Almaini O. et al., 2003, \mnras, 338, 303 
%
\bibitem[]{} 
Antonucci R., 1993, \araa, 31, 473 
%
\bibitem[]{} 
Aravena M. et al., 2008, \aap, 491, 173
%
\bibitem[]{} % paper I
Baldi A., Molendi S., Comastri A., Fiore F., Matt G., Vignali C., 2002, \apj, 564, 190
%
\bibitem[]{}
Berta S., Fritz J., Franceschini A., Bressan A., Pernechele C., 2003, \aap, 403, 119 
%
\bibitem[]{}
Blain A.~W., Smail I., Ivison R.~J., Kneib J.-P., Frayer D.~T., 2002, \physrep, 369, 111 
%
\bibitem[]{}
Cardelli J.~A., Clayton G.~C., Mathis J.~S., 1989, \apj, 345, 245 
%
\bibitem[]{}
Cavaliere A., Vittorini V., 2000, \apj, 543, 599 
%
\bibitem[]{} 
Chabrier G., 2003, \pasp, 115, 763 
%
\bibitem[]{}
Chapman S.~C., Blain A.~W., Ivison R.~J., Smail I.~R., 2003, \nat, 422, 695 
%
\bibitem[]{}
Chapman S.~C., Blain A.~W., Smail I., Ivison R.~J., 2005, \apj, 622, 772 
%
\bibitem[]{}
Charmandaris V., Laurent O., Mirabel I.~F., Gallais P., Sauvage M., Vigroux L., 
Cesarsky C., Tran D., 1999, \apss, 266, 99 
%
\bibitem[]{} % paper XI
Civano F. et al., 2007, \aap, 476, 1223 
%
\bibitem[]{} % paper VIII
Cocchia F. et al., 2007, \aap, 466, 31 
%
\bibitem[]{} % paper II
Comastri A. et al., 2002, \apj, 571, 771 
%
\bibitem[]{} % proc. Armenia
Comastri A., Vignali C., Brusa M., 2002, on behalf of the Hellas and Hellas2XMM Consortia, 
in ``AGN Surveys'', ed. R.~F. Green, E.~Ye. Khachikian, D.~B. Sanders, 
IAU Colloq.~184, 284, 235 
%
\bibitem[]{} % CT review
Comastri A., 2004, in ``Supermassive Black Holes in the Distant Universe'', 
ed. A.J. Barger, Kluwer Academic Press, Vol.~308, p.~245
%
\bibitem[]{}
Coppin K. et al., 2008, \mnras, 389, 45 
%
\bibitem[]{}
Di Matteo T., Springel V., Hernquist L., 2005, \nat, 433, 604 
%
\bibitem[]{}
Dye S. et al., 2008, \mnras, 386, 1107 
%
\bibitem[]{}
Fabian A.~C., 1999, \mnras, 308, L39 
%
\bibitem[]{}
Fabian A.~C. et al., 2000, \mnras, 315, L8 
%
\bibitem[]{}
Fadda D. et al., 2006, \aj, 131, 2859 
%
\bibitem[]{}
Ferrarese L., Merritt D., 2000, ApJ, 539, L9 
%
\bibitem[]{} % paper IV
Fiore F. et al., 2003, \aap, 409, 79 
%
\bibitem[]{}
Fiore F. et al., 2009, \apj, in press (arXiv:0810.0720)
%
\bibitem[]{}
Fritz J., Franceschini A., Hatziminaoglou E., 2006, \mnras, 366, 767  (F06)
%
\bibitem[]{}
Gebhardt K. et al., 2000, \apj, 543, L5 
%
\bibitem[]{}
Gordon K.~D. et al., 2007, \pasp, 119, 1019 
%
\bibitem[]{}
Granato G.~L., De Zotti G., Silva L., Bressan A., Danese L., 2004, \apj, 600, 580 
%
\bibitem[]{}
Gruppioni C. et al., 2008, \apj, 684, 136
%
\bibitem[]{}
Hatziminaoglou E. et al., 2008, \mnras, 386, 1252
%
\bibitem[]{}
Holland W.~S. et al., 1999, \mnras, 303, 659 
%
\bibitem[]{}
H{\"o}nig S.~F., Beckert T., Ohnaka K., Weigelt G., 2006, \aap, 452, 459 
%
\bibitem[]{}
%A Unified, Merger-driven Model of the Origin of Starbursts, Quasars, the Cosmic X-Ray Background, 
%Supermassive Black Holes, and Galaxy Spheroids
Hopkins P.~F., Hernquist L., Cox T.~J., Di Matteo T., Robertson B., Springel V., 2006, \apjs, 163, 1 
%
\bibitem[]{}
Hughes D.~H. et al., 1998, \nat, 394, 241 
%
%\bibitem[]{}
%Kennicutt R.~C. Jr., 1998, \apj, 498, 541 
%
\bibitem[]{}
Lacy M. et al. 2004, \apjs, 154, 166 
%
\bibitem[]{}
La Franca F. et al., 2005, \apj, 635, 864 
%
\bibitem[]{}
Lamastra A., Perola G.~C., Matt G., 2006, \aap, 449, 551 
%
\bibitem[]{}
Magorrian J. et al., 1998, \aj, 115, 2285 
%
\bibitem[]{}
Makovoz D., Marleau F.~R., 2005, \pasp, 117, 1113 
%
\bibitem[]{}
Mainieri V. et al., 2005, \mnras, 356, 1571 
%
\bibitem[]{}
Maiolino R. et al., 2006, \aap, 445, 457 
%
\bibitem[]{}
Marconi A., Hunt L.~K., 2003, \apjl, 589, L21 
%
%\bibitem[]{}
%McLure R.~J., Dunlop J.~S., 2004, \mnras, 352, 1390 
%
\bibitem[]{}
Menci N., Fontana A., Giallongo E., Grazian A., Salimbeni S., 2006, \apj, 647, 753 
%
\bibitem[]{}
Menci N., Fiore F., Puccetti S., Cavaliere A., 2008, ApJ, 686, 219
%
%\bibitem[]{} % paper V
%Mignoli M. et al., 2004, \aap, 418, 827 
%
\bibitem[]{}
Nenkova M., Ivezi{\'c} {\v Z}., Elitzur M., 2002, \apjl, 570, L9 
%
\bibitem[]{}
Nenkova M., Sirocky M.~M., Ivezi{\'c} {\v Z}., Elitzur M., 2008a, \apj, 685, 147 
%
\bibitem[]{}
Nenkova M., Sirocky M.~M., Nikutta R., Ivezi{\'c} {\v Z}., Elitzur M., 2008b, \apj, 685, 160  
%
\bibitem[]{}
Omont A., Beelen A., Bertoldi F., Cox P., Carilli C.~L., Priddey R.~S., McMahon R.~G., Isaak K.~G., 2003, \aap, 398, 857 
%
\bibitem[]{}
Page M.~J., Stevens J.~A., Mittaz J.~P.~D., Carrera F.~J., 2001, \sci, 294, 2516 
%
\bibitem[]{}
Page M.~J., Stevens J.~A., Ivison R.~J., Carrera F.~J., 2004, \apj, 611, L85 
%
\bibitem[]{} 
Page M.~J., Carrera F.~J., Ebrero J., Stevens J.~A., Ivison R.~J., 2006, in Charmandaris V., Rigopoulou D., Kylafis N., eds, 
``Studying Galaxy Evolution with Spitzer and Herschel'', in press (astro-ph/0610229)
%
\bibitem[]{}
Papovich C. et al., 2004, \apjs, 154, 70 
%
\bibitem[]{} % paper VI
Perola G.~C. et al., 2004, \aap, 421, 491 
%
\bibitem[]{} 
Piconcelli E., Jimenez-Bail{\'o}n E., Guainazzi M., Schartel N., 
Rodr{\'{\i}}guez-Pascual P.M., Santos-Lle{\'o} M., 2005, \aap, 432, 15 
%
\bibitem[]{} 
Pier E.~A., Krolik J.~H., 1992, \apj, 401, 99 
%
\bibitem[]{} % paper X
Pozzi F. et al., 2007, \aap, 468, 603 
%
\bibitem[]{}
Renzini A., 2006, \araa, 44, 141 
%
\bibitem[]{}
Richards G.~T., Vanden Berk D.~E., Reichard T.~A., Hall P.~B., 
Schneider D.~P., SubbaRao M., Thakar A.~R., York, D.~G., 2002, \aj, 124, 1 
%
\bibitem[]{}
Salpeter E.~E., 1955, \apj, 121, 161 
%
\bibitem[]{}
Sanders D.~B., Soifer B.~T., Elias J.~H., Madore B.~F., 
Matthews K., Neugebauer G., Scoville N.~Z., 1988, \apj, 325, 74 
%
\bibitem[]{}
Schartmann M., Meisenheimer K., Camenzind M., Wolf S., Tristram K.~R.~W., Henning T., 2008, \aap, 482, 
%
\bibitem[]{}
Severgnini P. et al., 2000, \aap, 360, 457 
%
\bibitem[]{}
Shupe D.~L. et al., 2008, \aj, 135, 1050 
%
\bibitem[]{}
Silk J., Rees M.~J., 1998, \aap, 331, L1 
%
\bibitem[]{}
Silva L., Granato G.~L., Bressan A., Danese L., 1998, \apj, 509, 103 
%
\bibitem[]{}
Smail I., Scharf C.~A., Ivison R.~J., Stevens J.~A., Bower R.~G., Dunlop J.~S., 2003, \apj, 599, 86 
%
\bibitem[]{}
Smail I., Ivison R.~J., Blain A.~W., 1997, \apj, 490, L5 
%
\bibitem[]{} 
Spergel D.N. et al., 2003, \apjs, 148, 175 
%
\bibitem[]{}
Steffen A.~T., Strateva I., Brandt W.~N., Alexander D.~M., Koekemoer A.~M., Lehmer B.~D., 
Schneider D.~P., Vignali C., 2006, \aj, 131, 2826 
%
\bibitem[]{}
Stern D. et al., 2005, \apj, 631, 163 
%
\bibitem[]{}
Stevens J.~A. et al., 2003, \nat, 425, 264 
%
\bibitem[]{}
Stevens J.~A., Page M.~J., Ivison R.~J., Smail I., Carrera F.~J., 2004, \apj, 604, L17 
%
\bibitem[]{}
Stevens J.~A., Page M.~J., Ivison R.~J., Carrera F.~J., Mittaz J.~P.~D., Smail I., McHardy I.~M., 
2005, \mnras, 360, 610 
%
\bibitem[]{}
Stansberry J.~A. et al., 2007, \pasp, 119, 1038 
%
%\bibitem[]{}
%Takata T., Sekiguchi K., Smail I., Chapman S.~C., Geach J.~E., 
%Swinbank A.~M., Blain A., Ivison R.~J., 2006, \apj, 651, 713 
%
\bibitem[]{}
Vanden Berk D.~E. et al., 2001, \aj, 122, 549 
%
\bibitem[]{}
Vignali C., Brandt W.~N., Schneider D.~P., 2003, \aj, 125, 433 
%
\bibitem[]{}
Waskett T.~J. et al., 2003, \mnras, 341, 1217 
%
%%%%%%%%%%%%%%%%%%%%%
\end{thebibliography}
